\documentclass[12pt,preprint,showpacs]{revtex4}
\usepackage{graphicx}
\DeclareRobustCommand{\baselinestretch{2.2}}

\begin{document}

\title{Phase sensitive detection of dipole radiation in a fiber-based high
  numerical aperture optical system}

\author{A. N. Vamivakas, A. K. Swan and M. S. \"{U}nl\"{u}}

\address{Department of Electrical and Computer Engineering, Boston
  University, 8 St. Mary's St., Boston, Massachusetts 02215}

\author{M. Dogan and B. B. Goldberg}

\address{Department of Physics, Boston
  University, 590 Commonwealth Ave., Boston, Massachusetts 02215}

\author{E. R. Behringer}

\address{Department of Physics and Astronomy, Eastern Michigan
  University, Ypsilanti, Michigan 48197 }

\author{S. B. Ippolito\footnote{Research conducted while at
    Boston University}}

\address{IBM T. J. Watson Research Center, 1101 Kitchawan Rd., 11-141, Yorktown Heights, New York 10598}


\begin{abstract} We theoretically study the problem of detecting
  dipole radiation in an optical system of high numerical aperture in
  which the detector is sensitive to \textit{field amplitude}.
   In particular, we model the
  phase sensitive detector as a single-mode cylindrical optical fiber.
   We find that the maximum in collection efficiency of the dipole
  radiation does not coincide with the optimum resolution for the
  light gathering instrument. The
  calculated results are important for analyzing fiber-based confocal microscope
  performance in fluorescence and spectroscopic studies of single molecules and/or
  quantum dots.
\end{abstract}




\maketitle 


The confocal microscope is a ubiquitous tool for the optical study and
 characterization of single nanoscale objects.  The rejection of stray 
light from the optical
detector afforded by the confocal microscope, combined with its
three-dimensional resolution, makes it an ideal instrument for
 studying  physical systems with weak light
 emission properties \cite{novotnyhecht, inoue}.  The electromagnetic dipole is the canonical
 choice for modeling the radiative properties of most physical
 systems.  And, although the 
vector-field image of an
 electromagnetic dipole in a high numerical aperture confocal microscope has
 been known for some time \cite{sheppard2}, only recently have the light gathering
 properties of the instrument been studied.  Specifically, the collection
 efficiency function for a confocal microscope based on a hard-stop
 aperture  was
 defined and studied by Enderlein \cite{enderlein}.   Such a confocal
 microscope is sensitive to field intensity and the detected optical
 power is obtained by integrating the component of the dipole
 image field 
Poynting vector that is perpendicular to the hard-stop aperture over the
 aperture area.  

Confocal microscopes based instead on
 optical fiber apertures have also been investigated.  The image forming
 properties of both coherent\cite{gu1} and
 incoherent\cite{gan1} fiber-based confocal microscopes, as well as the light
 gathering properties\cite{gu2} of the microscope with a reflecting object have all been
 examined assuming the paraxial approximation to scalar diffraction
 theory.   Since high numerical aperture fiber-based confocal microscopes are routinely
 employed in the study of silicon integrated circuits \cite{ippolito2}, single semiconductor quantum dots
 \cite{liu2} and other nanoscale light emitters,
 it is of great practical
 interest to understand the
 light collection properties of the fiber-based instrument.            
Here, we will extend the previous 
 studies by using the angular spectrum representation (ASR)\cite{wolf,richards,novotnyhecht} to
 study the coupling of dipole radiation into a single-mode optical
 fiber \cite{confocal}.

\begin{figure}
\begin{center}
\includegraphics[width=5in]{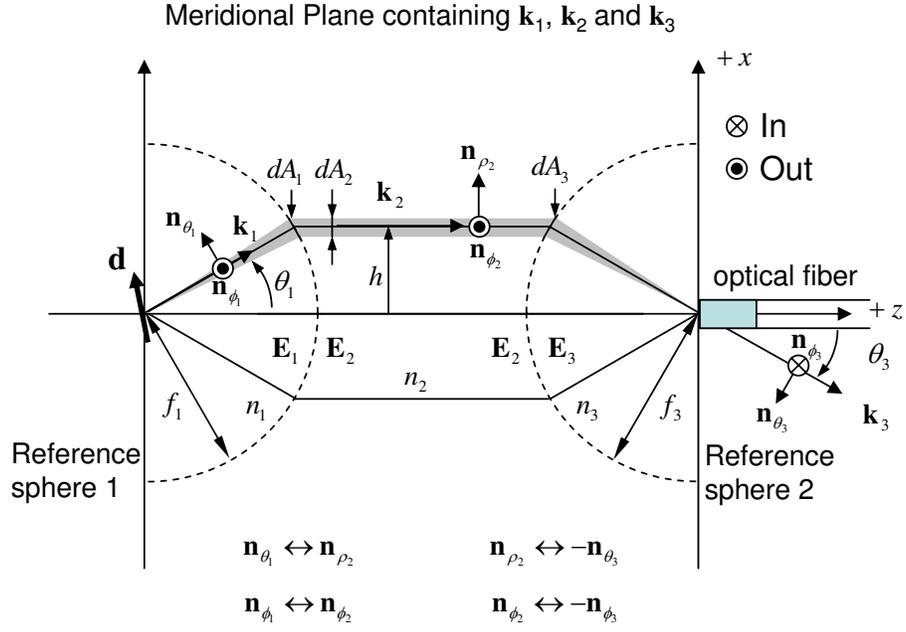}
\caption{The optical system geometry used to image an arbitrarily
  oriented dipole $\vec{\mathbf{d}}$.  The phase sensitive detector,
 an optical fiber, is situated in the image space of the 
microscope.} 
\label{Fig1}
\end{center}
\end{figure}

For the calculation below, we assume the optical system illustrated in
Fig. \ref{Fig1} is aplanatic.
 In what follows, we refer to reference
sphere 1 as the collection objective and reference sphere 2 as
the focusing objective.  Initially, we assume the dipole $\vec{\mathbf{d}}$ is placed at the 
Gaussian focus of the collection objective.  The cylindrical optical
fiber facet is assumed to be positioned such that it is coaxial with the optical system
axis (in Fig. \ref{Fig1}) and its flat face is parallel with the
focal plane of the focusing objective.  We define the relevant angles
and unit vectors in Fig. \ref{Fig1} as follows: 

\begin{equation}
 \label{eq:UnitVectors}
\begin{array}{c}
\hat{\mathbf{n}}_{\phi_1}=-\hat{\mathbf{n}}_{\phi_3}
 = -\sin{\phi_1}\hat{\mathbf{n}}_x + \cos{\phi_1}\hat{\mathbf{n}}_{y}\\
\hat{\mathbf{n}}_{\theta_1} =
-\hat{\mathbf{n}}_{\theta_3}
=
  \cos{\theta_1}\cos{\phi_1}\hat{\mathbf{n}}_x+\cos{\theta_1}\sin{\phi_1}\hat{\mathbf{n}}_{y}-
  \sin{\theta_1}\hat{\mathbf{n}}_{z}
\end{array}
\end{equation}

\noindent where we define the spherical coordinates $(\theta_1,\phi_1)$
($(\theta_3,\phi_3)$) in the object space (image
space) to
describe the orientation of the wavevector
$\vec{\mathbf{k}}_1$ ($\vec{\mathbf{k}}_3$),  and ensure that in each section of the
optical system all coordinate systems are right-handed.  In addition, 
the sine condition relates the polar angles in
the object and image space as $f_1\sin{\theta_1}=f_3\sin{\theta_3}$
where we have introduced the focal length $f_1$ ($f_3$)
for the collection (focusing) objective.
The geometry implies the azimuthal angles are related according to
$\phi_1=\phi_3-\pi$.

To calculate the vector-wave-optics image of the dipole, we employ the
ASR and express the image dipole
field as

\begin{eqnarray}
\label{eq:EASR}
 \vec{\mathbf{E}}_{3}(\rho_3,\varphi_3,z_3)
 =\frac{\omega^2}{\epsilon_o c^2}\frac{i\tilde{M}k_3 e^{if_1(k_1 - k_3\tilde{M})}}{8\pi^2}
\int_0^{\theta^{max}_3} \int_0^{2\pi}  d\theta_3 \,
 d\phi_3 \, \sqrt{\frac{n_1\cos\theta_3}{n_3\cos\theta_1}}\sin\theta_3 
 \nonumber \\
\times \biggl\{ -\hat{\mathbf{n}}_{\theta_3}[
 \hat{\mathbf{n}}_{\theta_3}\cdot
\vec{\mathbf{E}}_{d}(\theta_3,\phi_3)]
-\hat{\mathbf{n}}_{\phi_3}[ \hat{\mathbf{n}}_{\phi_3}\cdot\vec{\mathbf{E}}_{d}(\theta_3,\phi_3)]
\biggr\}
e^{ik_3(\rho_3\sin\theta_3\cos(\phi_3-\varphi_3)+z_3\cos\theta_3)}
\end{eqnarray}          

\noindent where we have defined the focal length ratio
$\tilde{M}=(f_3/f_1)$ (the magnification $M$ of the optical system
relates to $\tilde{M}$ as $M=\tilde{M}n_1/n_3$), the
magnitude of the wavevectors $k_i=\vert \vec{\mathbf{k}}_i\vert$, and integrate
over the polar angle $\theta_3$ and the azimuthal
angle $\phi_3$ in the image space.  To arrive at Eq. (\ref{eq:EASR}),
we map the vector electric far field of the dipole across the collection objective and
then across the focusing objective according to the unit vector mappings
defined in Fig. \ref{Fig1}.  We find it convenient to integrate over the object space
polar angle.  Introducing the identity
$d\theta_3 \sin\theta_3 = (\sin\theta_1\cos\theta_1 / \tilde{M}^2\cos\theta_3)d\theta_1$
into Eq. (\ref{eq:EASR}) and using Bessel function identities \cite{novotnyhecht}
to integrate over the azimuthal angle $\phi_3$ we find \cite{Sign}

\begin{eqnarray}
\label{eq:E3XFin}
\nonumber \vec{\mathbf{E}}^x_{3}(\rho_3,\varphi_3,z_3)
 =C(f_1,f_3)d_x
\left[\begin{array}{c}
 \tilde{I}_{d0}(\rho_3,z_3)+ \tilde{I}_{d2}(\rho_3,z_3)\cos2\varphi_3 \\
  \tilde{I}_{d2}(\rho_3,z_3)\sin2\varphi_3 \\
-2i\tilde{I}_{d1,2}(\rho_3,z_3)\cos\varphi_3   \end{array}\right],
\end{eqnarray}

\begin{eqnarray}
\label{eq:E3YFin}
\nonumber \vec{\mathbf{E}}^y_{3}(\rho_3,\varphi_3,z_3)
 =C(f_1,f_3)d_y
\left[\begin{array}{c}
\tilde{I}_{d2}(\rho_3,z_3)\sin2\varphi_3 \\
 \tilde{I}_{d0}(\rho_3,z_3)- \tilde{I}_{d2}(\rho_3,z_3)\cos2\varphi_3   \\
-2i\tilde{I}_{d1,2}(\rho_3,z_3)\sin\varphi_3   \end{array}\right],
\end{eqnarray}

\noindent and
     
\begin{eqnarray}
\label{eq:E3fin}
\vec{\mathbf{E}}^z_{3}(\rho_3,\varphi_3,z_3)
 =C(f_1,f_3)d_z
\left[\begin{array}{c}
2i\tilde{I}_{d1}(\rho_3,z_3)\cos\varphi_3 \\
2i\tilde{I}_{d1}(\rho_3,z_3)\sin\varphi_3\   \\
-2\tilde{I}_{d0,2}(\rho_3,z_3)   \end{array}\right]
\end{eqnarray}

\noindent where 

\begin{eqnarray}
\label{eq:C}
C(f_1,f_3)
 =\frac{\omega^2}{\epsilon_o c^2}\frac{i k_3 e^{if_1(k_1 -
 k_3\tilde{M})}}{8\pi \tilde{M}}
\sqrt{\frac{n_1}{n_3}}
\end{eqnarray}

\noindent and we use the notation
$\vec{\mathbf{E}}^j_{3}(\rho_3,\varphi_3,z_3)$ for the image field of a
$j$-oriented dipole in the object space expressed in terms of
Cartesian unit vectors.  The integrals $\tilde{I}_{dn}(\rho_3,z_3)$ are
defined as

\begin{eqnarray}
\label{eq:Id0-2NOf}
\begin{array}{c}
\tilde{I}_{d0}(\rho_3,z_3)=
\int_0^{\theta^{max}_1}   d\theta_1e^{ik_3z_3g(\theta_1)} 
\sqrt{\frac{\cos\theta_1}{g(\theta_1)}}\sin\theta_1
 \biggr(1+\cos\theta_1g(\theta_1)\biggl) J_0 \, ,
\nonumber \\ \nonumber \\
\tilde{I}_{d1}(\rho_3,z_3)=
\int_0^{\theta^{max}_1}   d\theta_1e^{ik_3z_3g(\theta_1)} 
\sqrt{\cos\theta_1}\sin^2\theta_1
  J_1 \, ,
\nonumber \\ \nonumber \\
 \tilde{I}_{d2}(\rho_3,z_3)=
\int_0^{\theta^{max}_1}   d\theta_1e^{ik_3zg(\theta_1)} 
    \sqrt{\frac{\cos\theta_1}{g(\theta_1)}}\sin\theta_1
 \biggr(1-\cos\theta_1g(\theta_1)\biggl)J_2 \, ,
\nonumber \\ \nonumber \\
 \tilde{I}_{d0,2}(\rho_3,z_3)=
\int_0^{\theta^{max}_1}   d\theta_1e^{ik_3z_3g(\theta_1)} 
    \sqrt{\frac{\cos\theta_1}{g(\theta_1)}}\frac{\sin^3\theta_1}{\tilde{M}^2}
 J_0 \, ,
\end{array}
\end{eqnarray}

\noindent and

\begin{eqnarray}
\label{eq:Iint}
\begin{array}{c}
\tilde{I}_{d1,2}(\rho_3,z_3)=
\int_0^{\theta^{max}_1}   d\theta_1e^{ik_3z_3g(\theta_1)} 
    \sqrt{\frac{\cos\theta_1}{g(\theta_1)}}\frac{\cos\theta_1\sin^2\theta_1}{\tilde{M}^2}
    J_1 
\end{array}
\end{eqnarray}

\noindent where $g(\theta_1) = \sqrt{1-(\sin\theta_1/\tilde{M})^2}$,
the numerical aperture ($NA_1$) in the object space defines $\theta^{max}_1$ as
$NA_1 = n_1\sin\theta^{max}_1$ and
$J_{m}$ are order $m=0,1,2$ ordinary Bessel functions with argument $(k_3\rho_3/\tilde{M})\sin\theta_1$.
  Equations (\ref{eq:E3fin}) - (\ref{eq:Iint}) assume the dipole is
situated at the Gaussian focus of the collection objective.  To
express the image of a displaced dipole located at $(\rho_o,\phi_o,z_o$), we use the
imaging property of the optical system and introduce
$\rho_{new}=\rho_{3}+M\rho_o$, $\varphi_3 = \varphi_o$ and $z_{new} =
z_3 + z_oM^2(n_3/n_1)$
into 
Eqs. (\ref{eq:E3fin}) - (\ref{eq:Iint}) where $M$ is
the optical system magnification.

We model the case when the phase sensitive detector of the dipole field
is a single-mode cylindrical optical fiber situated in the
image space of the optical system.  We define the collection
efficiency $\eta(\vec{\mathbf{r}}_o,\vec{\mathbf{d}};\tilde{M})$ of the optical fiber as

\begin{eqnarray}
\label{eq:CEFF}
\eta(\vec{\mathbf{r}}_o,\vec{\mathbf{d}};\tilde{M})= \frac{\vert\int\int\vec{\mathbf{E}}^{\ast}_{3}
(\vec{\mathbf{r}}_3;\vec{\mathbf{r}}_o,\vec{\mathbf{d}})\cdot
\vec{\mathbf{E}}^{j}_{lm}(\vec{\mathbf{r}}_3)\,dA_3\vert^2}
{ \int\int\vert\vec{\mathbf{E}}_{3}
(\vec{\mathbf{r}}_3;\vec{\mathbf{r}}_o=0,\vec{\mathbf{d}})\vert^2\, dA_3
 \int\int\vert\vec{\mathbf{E}}^{j}_{lm}(\vec{\mathbf{r}}_3)\vert^2 \,dA_3}
\end{eqnarray}

\noindent where we make explicit the dependence of
$\eta(\vec{\mathbf{r}}_o,\vec{\mathbf{d}};\tilde{M})$ 
on the dipole location $\vec{\mathbf{r}}_o$ and orientation
$\vec{\mathbf{d}}$ in the object space, and on
the objective focal length ratio $\tilde{M}$ of the optical system illustrated in Fig. \ref{Fig1}
(we condense notation by introducing
$\vec{\mathbf{r}}_3=(\rho_3,\varphi_3,z_3)$).  We point out the
collection efficiency, as defined in Eq. (\ref{eq:CEFF}), depends on
the overlap of the
dipole image field amplitude with the fiber mode profile and not on
the intensity of the dipole image field.  For
the single-mode optical fiber we make the weakly guiding
approximation\cite{gloge} and assume the cladding refractive index, $n_{cl}$, is
nearly equal to the core refractive index, $n_{co}$.  The
utility of the weakly guiding approximation is that the propagating mode
solutions for the fiber,
$\vec{\mathbf{E}}^{j}_{lm}(\vec{\mathbf{r}}_3)$, are linearly
polarized (along the direction indexed by $j$).  For each
propagating solution, characterized by propagation constant $\beta$,
there exist two orthogonal, linearly polarized modes
typically referred to as the $\mathbf{LP}_{lm}$ modes.
Specifically, for a fiber with core radius $a$, the single-mode fiber electric field solutions are\cite{buck}  

\begin{eqnarray}
\label{eq:ExSMF}
\vec{\mathbf{E}}^{x}_{01}(\vec{\mathbf{r}},t) = \left \{
\begin{array}{c c}
\sqrt{\frac{2Z}{\pi a^2}\frac{1}{\frac{J^2_{1}(u)V^2}{w^2}}}
J_0(\frac{ur}{a}) e^{i\beta z} \, \mathbf{\hat{n}}_{x}    &
r\leq a \\
\sqrt{\frac{2Z}{\pi a^2}\frac{1}{\frac{J^2_{1}(u)V^2}{w^2}}}
\frac{J_0(u)}{K_0(w)}K_0(\frac{wr}{a}) e^{i\beta z} \, \mathbf{\hat{n}}_{x}    &
r\geq a \\
\end{array}
\right.
\end{eqnarray}

\noindent where $u=a\sqrt{n^2_{co}k^2_o-\beta^2}$ and
  $w=a\sqrt{\beta^2 - n^2_{cl}k^2_o}$ are the transverse wavenumbers
  in the fiber core and cladding, $V^2 = u^2+w^2=ak_o\sqrt{n^2_{co}-n^2_{cl}}$ 
  is the fiber $V$-parameter, $K_l$ is the order $l$ modified Bessel
  function of the second kind, $Z$ is the characteristic impedance of
  the fiber core and the solution for the orthogonally polarized
  solution is obtained by interchanging $x$ with $y$ in Eq. (\ref{eq:ExSMF}). 

\begin{figure}
\begin{center}
\includegraphics[width=5in]{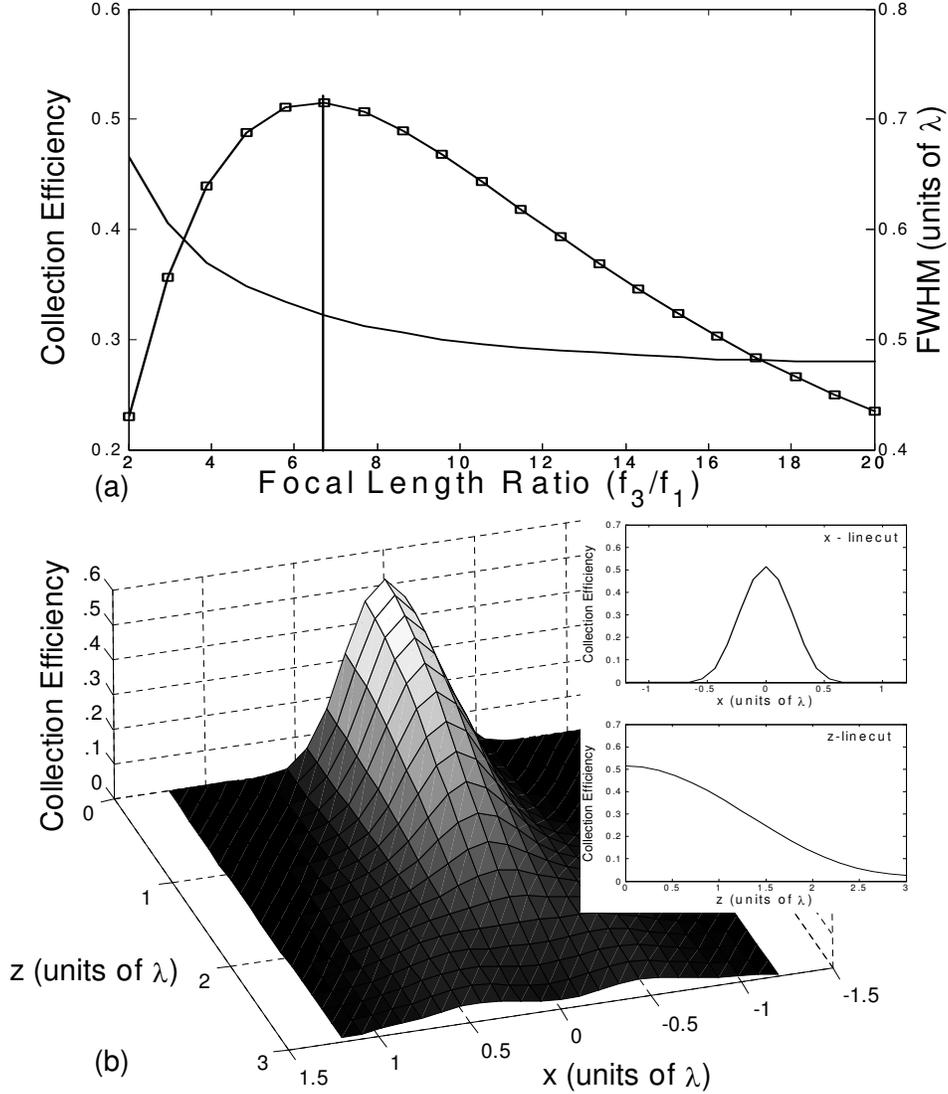}
\caption{(a) The collection efficiency defined in Eq. (\ref{eq:CEFF})
  and the full width at half maximum (FWHM) of the linecut
  $\eta(x_o=x,z_o=0)$ as a function of $\tilde{M}=f_3/f_1$.  The
  curves make apparent the compromise between collection efficiency and
  optical resolution.  The solid
  vertical line is the collection efficiency and FWHM for $\tilde{M} =
  NA_1/(n_1NA_3)$ where we use $NA_3=0.13$ for the assumed single mode fiber. 
(b) The collection
  efficiency 
defined in Eq. (\ref{eq:CEFF})
  as the dipole is
  displaced in the object space of the microscope fixing $M=9.31$.  The inset of (b) shows
  linecuts along $(x_o=x,z_o=0)$ and $(x_o=0,z_o=z)$.  For
  both (a) and (b)
  $n_1=1.33$, $n_3=1$, $a=0.5\lambda$, $V=1.03$, and the collection objective
  $NA_1=1.2$.}
\label{Fig2}
\end{center}
\end{figure}

Next, we apply the previous formalism to study the collection efficiency of
a fiber-based confocal microscope.  First, we position the dipole in a
region of refractive index $n_1=1.33$ at
the focus (equal to the coordinate origin) of a $NA_1 =1.2\,$ collection
objective and 
calculate the collection
efficiency $\eta(\vec{\mathbf{r}}_o=0;\tilde{M})$, averaged over a uniform
distribution of dipole orientations in the object space, as a function of $\tilde{M}$.
In addition, the single-mode fiber core radius is fixed to
$0.5\lambda$ ($\lambda$ is the wavelength of the dipole radiation) and the fiber $V$-parameter is equal to 1.03.
For the case of the two linearly polarized fiber modes, the collection
efficiency is expressible as an incoherent sum of the contribution
from each fiber polarization mode $\eta = \eta^x + \eta^y$ (we assume
the modes are linearly polarized along the $x$ and $y$ directions).
The result is the solid black line decorated with squares plotted in
Fig. \ref{Fig2}(a), showing that the maximum collection efficiency is obtained when
the ratio of the two objective focal lengths $\tilde{M}=f_1/f_3=7$
(corresponding to an optical system magnification of
$M=9.31$).  At this focal length ratio, we calculate a coupling
efficiency of approximately 
fifty-one percent.
From our definition of Eq. (\ref{eq:CEFF}), fifty-one percent of the
dipole radiation that enters the microscope image space is
coupled into the single-mode optical fiber.

Fixing the magnification to $M=9.31$, and
keeping $NA_1$ of the collection objective equal to 1.2, we calculate
collection efficiency $\eta(\rho_o=x,\phi_o=0,z_o=z;\tilde{M}=7)$ when the dipole
is displaced in the object space. The results are
presented in Fig. \ref{Fig2}(b).  The inset of Fig. \ref{Fig2}(b) displays linecuts
along $x$ $(z_o=0)$ and $z$  $(x_o=0)$.  We find a full width at
half maximum (FWHM) of approximately $0.522\lambda$ along the $x$-direction and
approximately $2.92\lambda$ along the axis of the microscope.  The product of
these numbers provides us with a rough estimate of the dipole
radiation collection volume (three-dimensional optical resolution) for
this
fiber-based confocal microscope.  In this case the number is approximately $0.795\lambda^3$.
We also studied the transverse resolution (along the $x$-direction) of
the optical system by calculating the FWHM as the
focal length ratio was varied around the value that resulted in
maximum collection efficiency.  The solid black line in
Fig. \ref{Fig2}(a) is the result of the calculation. We find that the
minimum of the FWHM (the optimal resolution)
does not coincide with the
 maximum of collection
efficiency.  At the focal length ratio $\tilde{M}$ that maximizes the
collection efficiency the transverse
 resolution is approximately nine
percent larger than the optimal transverse resolution. Finally, for comparison,
the solid vertical line in Fig. \ref{Fig2}(a) is both the collection
efficiency and transverse resolution when $\tilde{M}=NA_1/(n_1NA_3)$
where $NA_3=0.13$ for the assumed single-mode fiber.  By choosing the focal length ratio to match the refractive
index-scaled numerical
aperture ratio, the ability of the resulting optical system to collect
radiation from the dipole is maximized.

In summary, for a set of fixed optical system constraints, 
we find that there is a particular value of another system parameter
that optimizes the
overlap of the conjugated dipole image field amplitude with the fiber
mode profile and maximizes the collection efficiency as defined in
Eq. (\ref{eq:CEFF}).   In the example here, for fixed collection
objective numerical aperture
 and single-mode fiber characteristics, there is a particular value of
 the objective focal length ratio $\tilde{M}$ that maximizes the
 collection efficiency $\eta$.
 However, Fig. \ref{Fig2}(a) makes clear that in
constructing a fiber-based confocal microscope there is a compromise
between instrument collection efficiency and optical resolution.  It
is important in system design to determine which figure of merit,
collection efficiency or resolution, is most important.

\section*{Acknowledgments}
This work was supported by Air Force Office of Scientific Research
under Grant No. MURI F-49620-03-1-0379, by NSF under Grant No. NIRT
ECS-0210752 and a Boston University SPRInG grant.  The authors thank
Lukas Novotny 
for his helpful discussions on the angular
spectrum representation.


\begin{thebibliography}{10}
\bibitem{confocal}
In our analysis we assume a point emitter and point detector so we refer to the
  optical system as a confocal microscope. However, our results can be applied
  to other optical systems since there is no assumption on the mechanism for
  dipole excitation.
\bibitem{Sign}
Our results for the image of the dipole differ from those of Novotny and Hecht \cite{novotnyhecht}
by a minus sign for the x and y field components of an x-oriented and
y-oriented dipole.  The sign difference does not effect physically
important quantities such as energy, power flux or intensity.
\end{thebibliography}

\newpage

\end{document}